\documentclass[12pt]{article}

\usepackage[auth-sc,affil-sl]{authblk}
\usepackage[margin=0.98in]{geometry}
\usepackage{graphicx}
\usepackage[]{natbib}
\usepackage{abstract}
\usepackage{hyperref}

\newcommand{\qu}[1]{``#1''}

\title{Starvation of Cancer via Induced Ketogenesis and Severe Hypoglycemia}
\author[1]{Adam Kapelner\thanks{Electronic address: \texttt{kapelner@qc.cuny.edu}}}
\author[2]{Matthew Vorsanger\thanks{Electronic address: \texttt{matthew.vorsanger@nyumc.org}}}
\affil[1]{\normalsize{Department of Mathematics, Queens College, City University of New York}}
\affil[2]{Department of Cardiology, New York University}

\date{}

\begin{document}
\maketitle

%\begin{center}
%\inblue{In preparation for submission to \textit{Medical Hypotheses}}
%\end{center}

\begin{abstract}
Neoplasms are highly dependent on glucose as their substrate for energy production and are generally not able to catabolize other fuel sources such as ketones and fatty acids. Thus, removing access to glucose has the potential to starve cancer cells and induce apoptosis. Unfortunately, other body tissues are also dependent on glucose for energy under normal conditions. However, in human starvation (or in the setting of diet-induced ketogenesis), the body \qu{keto-adapts} and glucose requirements of most tissues drop to almost nil. Exceptions include the central nervous system (CNS) and various other tissues which have a small but obligatory requirement of glucose. Our hypothesized treatment takes keto-adaptation as a prerequisite. We then propose the induction of severe hypoglycemia by depressing gluconeogenesis while administering glucose to the brain. Although severe hypoglycemia normally produces adverse effects such as seizure and coma, it is relatively safe following keto-adaptation. We hypothesize that our therapeutic hypoglycemia treatment has potential to rapidly induce tumor cell necrosis.
\end{abstract}

\section{Introduction}\label{sec:introduction}

In 1924, Otto Heinrich Warburg's Nobel-prize winning research demonstrated that neoplastic cells rely on anaerobic glycolysis for their metabolic needs. Almost a century later, the results and implications of his findings are still hotly debated. It is now accepted that most tumors derive the vast majority of their energy from glucose. This major weakness is an attractive target for therapeutic intervention. However, basic physiology dictates that normal cells would also be starved by severe hypoglycemia. As such, targeting cancer by exploiting this metabolic weakness has not been proven fruitful to date.\footnote
{
There has been renewed interest with the success of positron emission tomography, a diagnostic tool which detects glucose uptake by making use of 18-fluorine 2-deoxyglucose.
}
In fact, cancer treatment has moved away from broadly cytotoxic modalities towards highly targeted therapies \citep{Aggarwal2010}. Treatments of choice include monoclonal antibodies, tyrosine kinase inhibitors, and induction of specific immune responses as well as others.

In this paper, we reexamine the possibility of therapeutic hypoglycemia as an antineoplastic treatment while simultaneously delivering glucose to the body tissues that require it. Our main hypothesis is that the alterations in metabolism that occur in the starvation state, namely neurologic adaptation to hypoglycemia, provide a means of safely inducing hypoglycemia. Of course, this proposal is not mutually exclusive with other standard treatments such as radiation, chemotherapy and dietary supplementation.

The outline of the paper is as follows. Section~\ref{sec:cancer} provides background on cancer cell metabolism and reviews the known cytotoxic effects of hypoglycemia on cancerous cells. Here, we justify the assumptions needed for our hypothesized treatment from the literature. The next two subsections are prerequisites for understanding the implementation of our hypothesized treatment. Section~\ref{subsec:starvation} discusses the body's theoretical minimum glucose requirements during starvation as well as the concept of tissue keto-adaptation. Section~\ref{subsec:keto} discusses ketogenic dieting as a viable alternative to starvation while retaining bodywide minimal glucose needs. We outline our severe hypoglycemia therapy in detail in Section~\ref{subsec:treatment}. Section~\ref{sec:discussion} concludes and discusses further extensions as well as potential concerns.

\section{Background}\label{sec:cancer}

Cancer is the result of multiple changes in the delicate balance of cell function. The root cause of neoplastic transformation is still under serious debate. There are many theories: cancer is primarily a genetic disease \citep{Vogelstein2013}, cancer is a disease of aberrant metabolism due to dysfunctional respiration in malfunctioning mitochondria \citep{Seyfried2010}, cancer is due to overproduction of reactive oxygen species \citep{Oleksyszyn2014}, as well as others.

Regardless of the cause, the majority of neoplastic cells feature an aberrant glucose meta-bolism first discovered by \citet{Warburg1924} and now known as the \qu{Warburg Effect.} Normal cells metabolize glucose in two steps. First, glucose is broken down to create pyruvate and a small amount of adenosine triphosphate (ATP) in the cytosol (glycolysis). Second, pyruvate is shuttled into mitochondria where it is fully oxidized to water and carbon dioxide in the Krebs cycle. Even with enough oxygen, cancer cells do not seem to advance beyond cytosolic glycolysis to employ the vastly more efficient (by a factor of 17 : 1) oxidative phosphorylation. Instead, they divert their pyruvate to produce lactic acid. This is oxymoronically known as \qu{aerobic glycolysis;} cancer cells ferment even in conditions of normoxia. A potential cause of this aberrant metabolism is that cancer cells only have a fraction of the mitochondria available to normal cells and these mitochondria are structurally defective (see a review by \citealp{Pedersen1977} and see \citealp{Elliott2011} for an electron microscopy study of mitochondria from breast carcinoma). Others say the cause is damaged glycolysis regulation \citep{Koppenol2011}. Regardless, \qu{all roads to the origin and progression of cancer pass through the mitochondria} and the hallmark of cancer is malfunction in oxidative phosphorylation to varying degrees \citep{Seyfried2014}.

This implies that cancer cells \textit{do not have access} to non-glycolytic fuels that demand full oxidative combustion in the Krebs cycle, namely fatty acids and the ketone bodies beta-hydroyxbutyrate ($\beta$-OHB) and acetoacetate \citep{Holm1995, Sawai2004}, two substrates that are key players in our hypothesized treatment. \citet{Chang2013a} demonstrated that the glioma cells in their investigation lack enzymes that break down ketones. Such evidence may explain why introduction of healthy mitochondria into cancer cell cytoplasm halts carcinogenesis (see \citealp[Chapter 11]{Seyfried2012} for a review of these studies).

The relevant point to our hypothesis is that glucose is the predominant energy substrate for most cancers \citep{Gullino1967} with glycolytic rates 8-200 times higher than normal tissues \citep[Chapter 5]{Phelps2004} while producing about 10\% more ATP than normal cells \citep{Koppenol2011}. Cancer has a large fuel requirement due to its high proliferation rate and large need for antioxidants. In addition, high glucose consumption is necessary for intermediaries required in biosynthetic pathways: ribose for nucleotides; glycerol, pyruvate and citrate for lipids; amino acids; and nicotinamide adenine dinucleotide phosphate (NADPH) via the pentose phosphate pathway \citep{DeBerardinis2008}. 

Glycolysis being the exclusive source of energy in cancer is hotly debated and may not be true of all cancers throughout their lifespan; tumors may have varied bioenergetic profiles. They may be glycolytic and/or oxidative \citep{Jelluma2006,Moreno-Sanchez2007,Martinez-Outschoorn2011,Jose2011,Carracedo2013}. For the purposes of this article, we address the majority of neoplasms which the consensus holds are exclusively glycolytic. We await future research that will elucidate when and in which cancer lines our hypothesized treatment is most applicable.

Thus, cancer should be particularly vulnerable to glucose / glycolytic substrate deprivation. Starving cancer from glucose is not a novel idea; proposals to treat cancer via serum hypoglycemia date back to an editorial in the \citet{NYT1887} and are abundant in the recent literature \citep{Woolf2014, Simone2013, Fine2012, Fine2008, Seyfried2008, Mavropoulos2006, Kim1978}. Another idea would be to inhibit glycolysis \citep[e.g.][]{Ganapathy-Kanniappan2010} but this is not the strategy we propose due to its more universal cytotoxicity.

Even though the idea of cancer starvation is commonplace, there have been no well-designed studies evaluating this premise. Why do they not exist? \citet{Simone2013} argues that the absence of clinical trials is due to unclear diet implementation protocols. Financial concerns are also a likely explanation. With the time and resources needed to produce a high-quality clinical trial, it is unlikely that therapeutic hypoglycemia would produce an adequate return-on-investment for a pharmaceutical company. At the time of this writing, there are some trials proposed (e.g. \citealp{TanShalaby2013} and the list found in Table 2 of \citealp{Simone2013}). There are many uncontrolled studies that are optimistic. For example, \citet{Zuccoli2010} details the account of a woman who had spontaneous remission (i.e. sudden disappearance of cancer) of glioblastoma multiforme after two months on a diet formulated to lower serum glucose. \citet{Niakan2010} concluded that over 1,000 similar spontaneous remission are most likely due to hypoglycemia and hypoxia (arguably a direct consequence of the hypoglycemia).

Pre-clinical animal studies demonstrate promising results by cutting cancer's nutrient supply \citep{Mukherjee2002, Mukherjee2004, Zhou2007, Otto2008, Mavropoulos2009, Shelton2010, Stafford2010, DeLorenzo2011, Sivananthan2013, Jiang2013}. However, there are some studies that are inconclusive \citep[e.g.][]{Masko2010}.

There are many creative in-vitro studies of glucose starvation. In \citet{Demetrakopoulos1978}, tumor cells were found to be highly sensitive to glucose deprivation. \citet{Spitz2000} found that glucose withdrawal induces cytotoxicity in transformed fibroblasts and colorectal cells which they argue is the result of oxidative stress. \citet{Buzzai2005} demonstrated that activation of the commonly expressed oncogene Akt prevents cancer cells from metabolizing non-glycolytic substrates.  \citet{Jelluma2006} performed the first study of glucose withdrawal on gliobastoma multiforme cells and found 90\% cell death in 24 hours. The death was not due to ATP depletion and could be rescued by a free radical scavenger, thereby lending credence to the oxidative stress theory. \citet{Aykin-Burns2009} demonstrated that glucose deprivation can induce cytotoxicity in human breast cancer cells and colon cancer cells in-vivo. They argue these results are due to cancer being unable to synthesize reducing agents such as NADPH and pyruvate which are needed \textit{en masse} for detoxification of high concentrations of free radicals due to mitochondrial abnormalities. \citet{Li2010a} observed growth inhibition and apoptosis in lung fibroblasts after glucose restriction and \citet{Priebe2011} demonstrated cell death in ovarian cancer. Also, \citet{Graham2012} demonstrated that glucose withdrawal slows the proliferation, induces hyperthermia, and possibly induces cancer cell death in four gliobastoma cell lines.

Our proposal is a natural extension of previous cancer starvation proposals. We propose a ketogenic diet which allows for the safe induction of \textit{severe} hypoglycemia via gluconeogenesis attenuation. The serum levels we propose are lower than have ever been proposed in the past and we do so in order to starve cancer cells rapidly. We now turn to detailing this therapy.

\section{Our Hypothesis}\label{sec:hypothesis}

The next two sections provide background material about human starvation and ketogenic dieting that is necessary for understanding our hypothesized treatment which we detail in Section~\ref{subsec:treatment}.

\subsection{Starvation: Minimizing the Body's Glucose Needs}\label{subsec:starvation}

The body's glucose requirements are minimized during periods of starvation, which constitutes a drastic transformation in the body's energy metabolism. Broadly speaking, starvation has five stages which are outlined in Table~\ref{tab:starvation_stages} below. 

\begin{table}[htp]
\centering
\begin{tabular}{lll}
Stage of & Physiological & Time \\
Starvation & Description & Period \\ \hline
1 & Gastrointestinal absorption & $< 1$ day \\
2 & Glycogenolysis & $< $ 2 days\\
3 & Gluconeogenesis & $\geq 2$ days \\
4 & Ketosis & $\geq 3$  days \\
5 (prolonged) & Decreased gluconeogenesis and increased & $\geq 14$ days  \\
   &  cerebral ketone consumption &
\end{tabular}
\caption{The five stages of starvation from the time of last ingestion (reprinted with slight modifications from \citealt[page 2]{Cahill1983}).}
\label{tab:starvation_stages}
\end{table}

Normally by the end of the first day, hepatic glycogen stores are depleted and the liver (and to a lesser extent, the kidneys) begin to produce glucose from pyruvate, glycerol and amino acids. The vast majority of tissues in the body (such as skeletal muscle, the heart, and most organs) are facultative in their choice of substrate for energy production. During this period of glucose scarcity, they cut their glucose metabolism and increase their metabolism of free fatty acids (FFAs). Starvation even at this early stage forces body tissues to switch from glycolysis to lipid oxidation for their energy needs. The \textit{ketone bodies} $\beta$-OHB, acetoacetate and acetone are always present in the blood, but after a few days of starvation they begin to be produced in large quantities in the liver. Most tissues of the body can metabolize ketones (with the exception of acetone).

Of special interest is stage 5 --- prolonged starvation of two weeks and longer. Cells have long since cut their uptake of glucose but now also cut their uptake of ketones. This change, coupled with increased blood-brain-barrier permeability of ketones \citep{Morris2005} we will call \qu{keto-adaptation.} During stage 5, most cells rely purely on FFAs for their metabolic requirements \citep{Robinson1980}. However, some tissues still retain a need for glucose even after keto-adaptation. The largest consumer is the brain and CNS, whose health is vital to the success of our hypothesis.\footnote
{
Note that erythrocytes, leukocytes and bone marrow also solely rely on glucose, but they only employ glycolysis. Thus, the lactate and pyruvate waste products are reconstituted into glucose via the Cori cycle in the liver using energy supplied by the breakdown of lipids. Therefore, these tissues have zero \textit{net} glucose consumption.
}

How much glucose does the brain need? A typical brain glucose requirement is approximately 144 g/day \citep[Figure 1]{Cahill1970}. This is a preferential utilization but not an \textit{obligatory} utilization. This point is crucial to our hypothesis. During keto-adaptation, the brain glucose requirement drops to approximately 44 g/day \citep[Figure 5]{Cahill1970}.\footnote
{
It is not known if the average brain can do with less than 44 g/day of glucose. We assume that if the body can get away with providing less, it would do so. Why? Each gram of glucose conserved is one gram of protein conserved. Protein sparing is essential during starvation since protein is man's precious machinery vital to cellular structure and function.  Depletion of these reserves can limit long-term survival even with massive lipid depots. This question can also be asked as follows: why are ketones unable to provide 100\% of the brain's energy requirements? The answer is unknown and we posit two theories. First, there may be incomplete $\beta$-OHB and acetoacetate transport to all cells of the CNS \citep{Morris2005}. Second, $\beta$-OHB and acetoacetate can only be converted to energy via oxidative phosphorylation in mitochondria; these organelles are scant in long and thin myelinated axons \citep[page 17]{Waxman1995}. Here, we posit that some of the brain's energy production must default to glycolysis in the axoplasm.
}

What supplies the energy that this difference of approximately 100 g/day of glucose previously supplied? The ketone bodies $\beta$-OHB and acetoacetate in a 6:1 proportion \citep[Table 5]{Owen1967}. Note that the brain does not metabolize FFAs \citep[Table 4]{Owen1967}.\footnote
{
This is a mystery as FFAs are able to cross the blood-brain-barrier and the brain is equipped with beta-oxidation enzymes for their combustion \citep{Morris2005}. 
}
In stage 5 starvation, ketone blood levels surge to 2.5-9.7 mmol/L which are in contrast to a negligible level of 0.01 mmol/L in non-starving subjects. The brain will readily use ketone bodies if they are available even in non-fasting periods, but under this condition ketone usage is negligible \citep{Sokoloff1973}. \citet{Morris2005} suspects that the brain's ketone metabolism was evolutionarily developed to ensure survival in times of glucose being scarce. \citet{Cahill1970} suspects that other body tissues slash their ketone consumption in favor of FFA's to ensure adequate substrate for the brain. Further, \citet{Cahill2003} argue that if metabolic adaptation to ketones did not exist, \textit{Homo sapiens} could not have developed such a large and advanced brain.

\begin{figure}[h]
\centering
\includegraphics[width=6.2in]{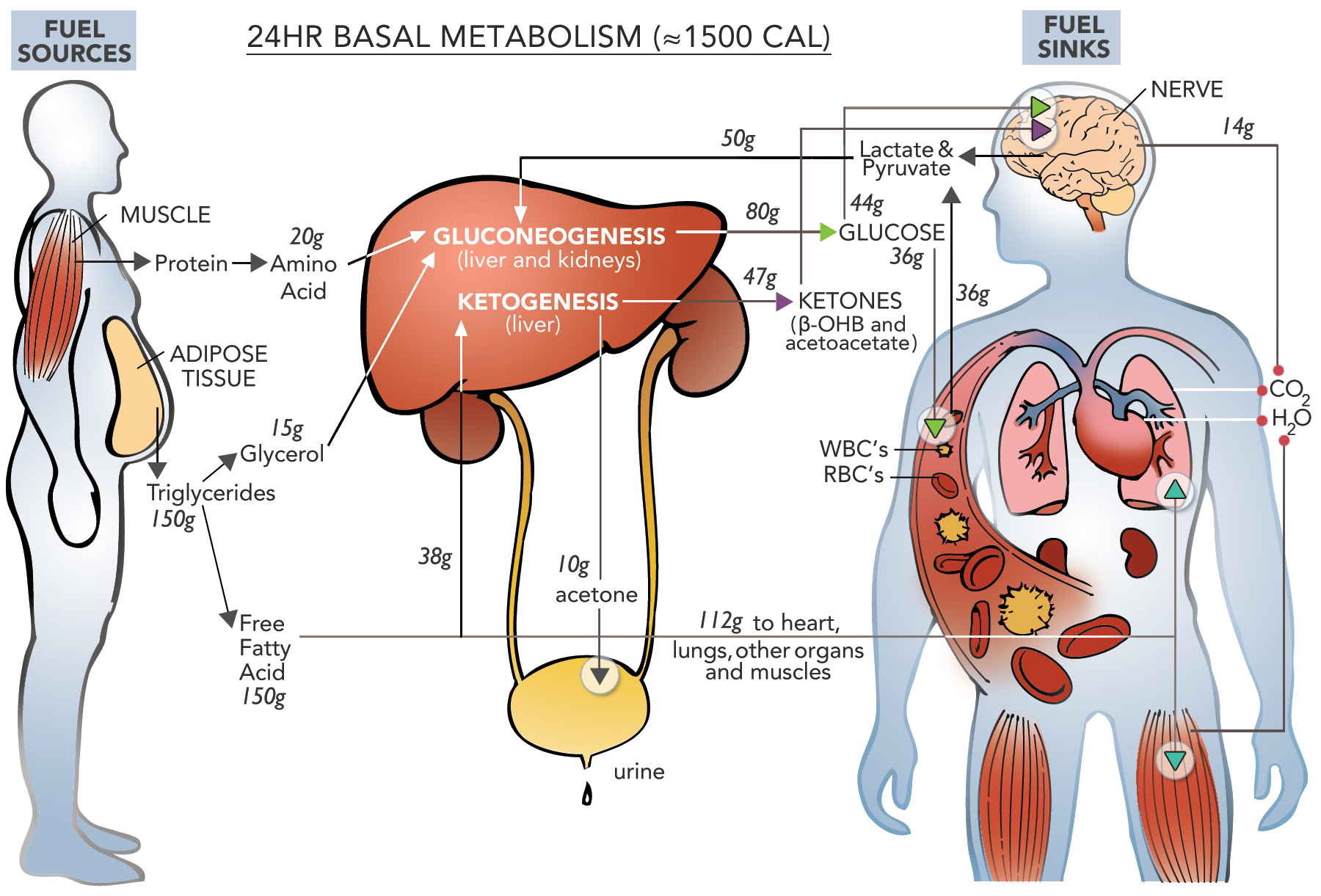}
\caption{Fuel sources and sinks in the daily metabolism of Stage 5 starvation assuming a 1500 calorie resting metabolism (inspired by \citealp[Figure 5]{Cahill1970}).}
\label{fig:starvation}
\end{figure}

The fuel metabolism of stage 5 starvation is illustrated in Figure~\ref{fig:starvation}. Of special interest is the inventory of the sources of gluconeogenesis that provide the 44 g/day CNS requirement: stored triacylglycerols (15g), amino acids taken from muscle tissue and to a lesser extent other tissues (20g), and pyruvate / lactate (14g), which is the end product of anaerobic glycolysis (thus about a third of the glucose during starvation burned by the CNS is recycled back into glucose via the Cori cycle).

Starvation can be a useful tool to minimize glucose needs in all but a few body tissues. Fasting, while relatively safe \citep{Stewart1973}, is not only unpleasant but leads to muscle wasting and weight loss, both of which are serious concerns in cancer patients. How do we minimize the body's glucose needs while avoiding these undesired effects? We turn to a solution in the next section.

\subsection{Keto-adaptation via the 4:1 Ketogenic Diet}\label{subsec:keto}

A means to emulate the metabolic conditions of starvation while still ingesting food is the same engineering problem faced in the 1920's by the expert childhood epileptologist Dr. Russel M. Wilder. He observed that short-term starvation was therapeutic for children with intractable epilepsy, but obviously impractical as a long-term solution. He innovated the \qu{ketogenic diet} (KD) that mimics the characteristic starvation metabolism we outlined in Section~\ref{subsec:starvation}. It is still used today to reduce seizures (for an historical account and a recent clinical trial, see \citealp{Neal2008}).

Wilder's KD is a high fat, low protein and very low carbohydrate diet \citep{McDonald1998}. To mimic starvation, strict macronutrient proportions are required.  In what is called the \qu{4:1 KD,} carbohydrate must be limited to 20 g/day (the amount the muscle tissue provides during prolonged starvation) and protein must be limited to an amount to offset normal (non-starvation) protein turnover, estimated at about 25 g/day for a 70kg person \citep{NRC1989}. Any additional protein has the potential to be gluconeogenic and thereby should be avoided. All other calories must be supplied by fats.\footnote
{
We detail the calculation here. 90\% of the calories are supplied by triglycerides which are esters of three fatty acid chains and one glycerol. 1/10th of fat calories are derived from the directly gluconeogenic glycerol. Thus, 81\% of dietary calories is supplied by fatty acids and 9\% + 10\% = 19\% by other substrates yielding a ratio of $81\% / 19\% \approx 4 : 1$. 
}
Commercial formulas for this diet are readily available.

Are the physiological markers in a prolonged 4:1 KD exactly the same as markers under prolonged starvation? Depending on the macronutrient proportions, serum glucose can be slightly higher and ketone levels slightly lower \citep{Phinney1983}. We assume that a 4:1 KD is equivalent to keto-adapted metabolism observed during prolonged starvation described in Section~\ref{subsec:starvation}: glucose uptake minimization, increased cerebral ketone uptake, and exclusive use of FFAs for energy production in every other tissue in the body. Thus, a KD can achieve starvation metabolism adaptation \textit{without} the unpleasant side effects of hunger and muscle loss. 

\citet{Simone2013} also reviews other benefits of the KD such as muscle sparing and possibly even muscle synthesis which can be beneficial for cancer patients with risk of cachexia. The KD is generally \qu{well-tolerated with minimal side effects such as constipation, salt loss, mild acidosis and increased incidence of kidney stones, when employed chronically} and has been demonstrated to be safe even in patients with advanced cancer \citep{Schmidt2011}. \citet{Kossoff2007} reports on the longest duration of a continuous KD: 21 years with virtually no side effects. There is also some evidence that ketones in their own right are directly toxic to cancer \citep{Magee1979, Skinner2009, Fine2009} and can improve the immune system's ability to target cancer \citep{Husain2013}. This can be an added benefit of the KD that can work in tandem with our hypoglycemia strategy.

Our hypothesized treatment begins with a cancer patient being administered a 4:1 KD for more than 14 days to allow for keto-adaptation. Then our proposed acute treatment is begun. We turn to implementation details now.

\subsection{Our Proposed Treatment}\label{subsec:treatment}

Section~\ref{sec:cancer} discussed previous literature that experimented with both starvation and the KD for cancer therapy via glucose starvation. We believe these studies did not reduce serum glucose to levels sufficient to starve the cancer. On a KD, \qu{plasma glucose concentration falls only mildly, remaining in the normal range in normal-weight individuals ... [thus] ... simple tumor glucose starvation is therefore unlikely in humans} \citep{Fine2008}. 

Serum glucose in prolonged starvation drops to about 65mg/dL \citep[Table 2]{Cahill1970} and the prolonged 4:1 KD can also result in a similar serum level, but it is unclear why further reduction in the blood glucose level should not be similarly well-tolerated after keto-adaptation.

In an adult, cerebral blood flow is approximately 750mL/min = 0.13 dL/s which implies 8.1mg/s of glucose can potentially be delivered to the brain. In Section~\ref{subsec:starvation}, we noted that the brain has an obligatory requirement of 44 g/day = 0.51 mg/s\footnote
{
This calculation is within 20\% of the brain arteriovenous glucose difference observed in starving subjects \citep[Table 5]{Owen1967}.
} 
which is only 6\% of the glucose available at a serum concentration of 65 mg/dL. Although experiments evaluating the glucose extraction ratio in human subjects are limited, animal studies suggest that the brain is able to extract 40\% of the serum glucose in the hypoglycemic state \citep{McCall1986}. Thus, we theorize that the body is being conservative. Serum glucose concentrations below the level where 0.51 mg/s can be actively transported into neurons for mere moments can result in CNS damage; 65 mg/dL is comfortably above this threshold.

We have evidence for this conjecture that keto-adapted subjects can survive on serum glucose much lower than 65mg/dL. Hypoglycemia has been reported in therapeutic fasting or starvation for over 100 years. \citet{Chakrabarty1948} investigated 407 starvation cases in the Bengal famine to find that 20 or so of them had chronic blood sugar below 40 mg/dL with no symptoms of hypoglycemia. \citet{Stewart1973} who supervised a world-record-setting 382-day therapeutic fast found that blood glucose became stable at 30 mg/dL, dropping intermittently to 20 mg/dL without symptoms. The most relevant evidence to our hypothesis is the observation of intentional acute hypoglycemia induction by \citet{Drenick1972} who keto-adapted obese patients via 45-60 days of starvation and then perfused insulin. Within the hour, they observed serum glucose as low as 9 mg/dL without acute hypoglycemic symptoms. Serum glucose this low would normally induce coma (which has even been observed with blood sugar as high as 40-49 mg/dL according to \citealp[Figure 3]{Ben-Ami1999}).

Our plan is simple. After keto-adaptation with the 4:1 KD, we propose to acutely attenuate the serum glucose concentration to severely hypoglycemic levels such as observed in \citet{Drenick1972} on a patient-by-patient basis. Since insulin is required by cancer cells and fuels their metabolism \citep{Iqbal2013} and also can impede ketogenesis, we choose not to induce hypoglycemia via insulin perfusion (as in \citealp{Drenick1972}), but instead propose the use of a gluconeogenesis inhibitor. 

One possible inhibitor is the diabetic drug metformin which inhibits gluconeogensis indirectly. Metformin may have other anti-cancer effects \citep[Figure 3]{Jalving2010} and there is certainly supporting epidemiological evidence for metformin's therapeutic effect in cancer \citep{BenSahra2010}. For these reasons, it has been proposed for use in a hypothesized treatment similar to ours (see \citealp{Oleksyszyn2011} and further in \citealp{Oleksyszyn2014}). However, metformin has many drawbacks. First, it is not a potent inducer of hypoglycemia in the doses that are normally administered. Metformin is also slow acting and not rapidly titratable. Additionally, the drug activates adenosine monophosphate activated protein kinase which may have other effects on metabolism under keto-adaptation \citep[see][Figure 2]{Jalving2010}. Lastly, supratherapeutic doses have been associated with fatal lactic acidosis.

Due to these shortcomings, we propose using a drug which is a pure liver-kidney gluconeogenesis inhibitor. Luckily, these drugs have been investigated by the diabetes community where high endogenous glucose production is considered one of the characteristics of the disease and is therefore a therapeutic target. 

There are many control points within gluconeogenesis. The unidirectional synthesis of fructose-6-phosphate from fructose-1,6-bisphospate is a prudent point to interrupt because it is removed from glycogenolysis and mitochondrial function (thus, it would have the fewest side effects). The synthesis is controlled via the rate-controlling enzyme fructose-1,6-bisphosphatase. Two inhibitors that have been clinically evaluated are named CS-917 and MB07803 \citep{Poelje2011}. They have good safety profiles and can flexibly attenuate gluconeogenesis to any desired level.

Returning to our discussion of the target level of hypoglycemia, we comment that we do not know the effective dose (ED) curves for serum glucose levels for tumor necrosis or the or lethal dose (LD) curves for human body death after keto-adaptation. It is possible that a 9 mg/dL glucose concentration may be too high to be toxic for the cancer. If this proves to be the case, further hypoglycemia would need to be safely induced for treatment efficacy. We turn to this now.

Since the brain is the only tissue with an obligatory glucose requirement, we opt for direct perfusion of glucose into the brain intra-arterially via the carotid arteries at the rate necessary to provide minimum brain glycolytic needs. Perfusion of glucose directly to the brain has been performed successfully in rats \citep{Borg1997} and baboons \citep{Conway1969}. Once the brain receives its required glucose, we propose an increase in the dose of gluconeogenic inhibitor drug to attenuate serum glucose to a negligible level which is safe enough for erythocytes, blood marrow and leukocytes to survive. The brain would perceive normoglycemia but everywhere outside of the brain the serum would be \textit{extremely} hypoglycemic. If necessary, we can also perfuse $\beta$-OHB in tandem if there is risk that the level of ketosis is not sufficiently high for CNS metabolic needs. This protocol would require extensive proof-of-concept work in animal models before any attempt to undertake study in human subjects.

The reader may wonder if this strategy would be effective against brain cancer since the brain is still receiving glucose. \citet{Seyfried2003} reports data supporting that such a treatment may be effective since glucose uptake in brain cancers drops with lower serum glucose and that the cancer cannot metabolize ketones, which may even be toxic to them.

How fast could this procedure induce tumor death? We echo \citet{Oleksyszyn2014}, \qu{we do not have any idea as to how deep hypoglycemia needs to be to trigger spontaneous remission.} All estimates are drawn from data from \textit{in-vitro} studies and thus are not generalizable to human prognoses. \citet[Figure 2]{Spitz2000} reports between 55\% and near 100\% necrosis rates in three cancer lines within 48hr after total glucose deprivation. Complete withdrawal is not possible in the human body for reasons explained above and our treatment could deliver minimum serum levels estimated at 5 mg/dL.

\section{Discussion}\label{sec:discussion}

We propose a therapy for starving cancer by inducing keto-adaptation followed by the induction of therapeutic hypoglycemia via gluconeogenesis inhibitor drugs. We then propose rescue cerebral glucose infusion through direct catheterization of the brains's blood supply. This would be followed by an increase of the gluconeogenic inhibitor drug to lower serum glucose even further in non-cerebral tissues. 

A main concern is that each person is a \qu{unique metabolic entity} \citep{Seyfried2014}. The calculations presented throughout this manuscript are meant to be illustrative of an average person. For each individual, personalized dosing and administration schedules will require iterative fine-tuning. 

Proper monitoring and its associated costs are a concern. Hypoglycemia is very risky but might not necessarily require the monitoring setting of an intensive care unit (patients with diabetic ketoacidosis are frequently treated with insulin infusion outside of the intensive care unit). Although the costs of the diet and the medications needed to induce hypoglycemia in our proposal might be far lower than the cost of novel chemotherapeutic agents, the cost of close monitoring may substantially erode this benefit.

There are several other concerns with this speculative treatment. Cancer is heterogeneous and few predictors are currently available as to an individual tumor's response to hypoglycemia. There is research suggesting that some cancer cells can make use of ketones and FFAs. Not all cancer lines proliferate rapidly and intense anabolism is not mandatory \citep{Jose2011}. Our hypothesized treatment would not work in such cancer lines. It is not known whether the proposed hypoglycemic levels can be sustained for long periods of time without symptoms. Starvation research suggests that hypoglycemia is safe in the long term, but these accounts documented serum levels many times above our proposed serum level and they were predominantly documented in obese subjects without cancer. An important consideration is whether keto-adaptation additionally protects against the neurohormal cardiovascular and electrophysiologic consequences of hypoglycemia \citep{Frier2011}. Additionally, glucose is not the sole nutritive substrate for cancer cells. For instance, glutamine can also serve as a major metabolite \citep{Yuneva2008}, as well as fermentation from other amino acids. We assume in our hypothesis that glucose is the necessary metabolite where other metabolites have neither the availability nor the capability of being transported in volumes necessary to serve as replacements.  Lastly, there are times when the body is in need of glucose such as during inflammation and infection which limits the setting of our hypothesized treatment. 

There is also a fair chance that our proposed treatment is more effective when provided in tandem with other therapies. For instance, \citet{Abdelwahab2012} demonstrated that brain cancer in mice was curtailed and did not recur with the simultaneous administration of the 4:1 KD and radiation therapy. \citet{Allen2013} demonstrated that the 4:1 KD enhances radio-chemo-therapy in mice models of lung cancer. See \citet{Klement2014} for a discussion of how the KD can act synergistically with radiation therapy. Also, \citet{Lee2012} demonstrated that fasting enhances chemotherapy. \citet{Poff2013} demonstrated that hyperbaric oxygen therapy is enhanced by the KD. 

Alternative adjuvant therapies may also be fruitful. For example, \citet{Majumdar2009} demonstrated that the combination of curcumin with resveratrol could be an effective therapy in colorectal cancer, \citet{Singh2004} demonstrated cancer apoptosis from wormwood extract and \citet{Tin2007} demonstrated that astralagus saponins can inhibit colorectal cancer cell proliferation. There are many other such supplements that can be administered simultaneously with our hypoglycemia treatment.

\subsection*{Acknowledgements}

We would like to thank Justin Bleich and Stephen Kapelner for helpful discussions and comments on this manuscript. We would like to thank Marie Le Pichon for illustrating Figure~\ref{fig:starvation}. Adam Kapelner acknowledges support from the National Science Foundation's Graduate Research Fellowship as well as support from the Simons Foundation Autism Research Initiative.

\small
\bibliographystyle{apalike}\bibliography{cancer_therapy}

\begin{thebibliography}{}

\bibitem[Abdelwahab et~al., 2012]{Abdelwahab2012}
Abdelwahab, M.~G., Fenton, K.~E., Preul, M.~C., Rho, J.~M., Lynch, A.,
  Stafford, P., and Scheck, A.~C. (2012).
\newblock {The ketogenic diet is an effective adjuvant to radiation therapy for
  the treatment of malignant glioma.}
\newblock {\em PloS one}, 7(5):e36197.

\bibitem[Aggarwal, 2010]{Aggarwal2010}
Aggarwal, S. (2010).
\newblock {Targeted cancer therapies.}
\newblock {\em Nature reviews. Drug discovery}, 9(6):427--428.

\bibitem[Allen et~al., 2013]{Allen2013}
Allen, B.~G., Bhatia, S.~K., Buatti, J.~M., Brandt, K.~E., Lindholm, K.~E.,
  Button, A.~M., Szweda, L.~I., Smith, B.~J., Spitz, D.~R., and Fath, M.~A.
  (2013).
\newblock {Ketogenic diets enhance oxidative stress and radio-chemo-therapy
  responses in lung cancer xenografts.}
\newblock {\em Clinical cancer research : an official journal of the American
  Association for Cancer Research}, 19(14):3905--13.

\bibitem[Aykin-Burns et~al., 2009]{Aykin-Burns2009}
Aykin-Burns, N., Ahmad, I.~M., Zhu, Y., Oberley, L.~W., and Spitz, D.~R.
  (2009).
\newblock {Increased levels of superoxide and H2O2 mediate the differential
  susceptibility of cancer cells versus normal cells to glucose deprivation.}
\newblock {\em The Biochemical journal}, 418(1):29--37.

\bibitem[Ben-Ami et~al., 1999]{Ben-Ami1999}
Ben-Ami, H., Nagachandran, P., Mendelson, A., and Edoute, Y. (1999).
\newblock {Drug-induced hypoglycemic coma in 102 diabetic patients.}
\newblock {\em Archives of internal medicine}, 159(3):281--4.

\bibitem[{Ben Sahra} et~al., 2010]{BenSahra2010}
{Ben Sahra}, I., {Le Marchand-Brustel}, Y., Tanti, J.-F., and Bost, F. (2010).
\newblock {Metformin in cancer therapy: a new perspective for an old
  antidiabetic drug?}
\newblock {\em Molecular cancer therapeutics}, 9(5):1092--1099.

\bibitem[Borg et~al., 1997]{Borg1997}
Borg, M.~A., Sherwin, R.~S., Borg, W.~P., Tamborlane, W.~V., and Shulman, G.~I.
  (1997).
\newblock {Local ventromedial hypothalamus glucose perfusion blocks
  counterregulation during systemic hypoglycemia in awake rats.}
\newblock {\em The Journal of clinical investigation}, 99(2):361--5.

\bibitem[Buzzai et~al., 2005]{Buzzai2005}
Buzzai, M., Bauer, D.~E., Jones, R.~G., Deberardinis, R.~J., Hatzivassiliou,
  G., Elstrom, R.~L., and Thompson, C.~B. (2005).
\newblock {The glucose dependence of Akt-transformed cells can be reversed by
  pharmacologic activation of fatty acid beta-oxidation.}
\newblock {\em Oncogene}, 24(26):4165--73.

\bibitem[Cahill and Veech, 2003]{Cahill2003}
Cahill, G. and Veech, R.~L. (2003).
\newblock {Ketoacids? Good Medicine?}
\newblock {\em Transactions of the american clinical and climatological
  association}, 114:149--163.

\bibitem[Cahill, 1970]{Cahill1970}
Cahill, G.~F. (1970).
\newblock {Starvation in man}.
\newblock {\em New England Journal of Medicine}, 282:668--675.

\bibitem[Cahill, 1983]{Cahill1983}
Cahill, G.~F. (1983).
\newblock {Starvation}.
\newblock {\em Transaction of the American Clinical and Climatological
  Association}, 94:1--21.

\bibitem[Carracedo et~al., 2013]{Carracedo2013}
Carracedo, A., Cantley, L.~C., and Pandolfi, P.~P. (2013).
\newblock {Cancer metabolism: fatty acid oxidation in the limelight.}
\newblock {\em Nature reviews. Cancer}, 13(4):227--232.

\bibitem[Chakrabarty, 1948]{Chakrabarty1948}
Chakrabarty, M.~L. (1948).
\newblock {Blood-sugar levels in slow starvation}.
\newblock {\em The Lancet}, 251(6503):596--597.

\bibitem[Chang et~al., 2013]{Chang2013a}
Chang, H.~T., Olson, L.~K., and Schwartz, K.~A. (2013).
\newblock {Ketolytic and glycolytic enzymatic expression profiles in malignant
  gliomas: implication for ketogenic diet therapy.}
\newblock {\em Nutrition \& metabolism}, 10(1):47.

\bibitem[Conway et~al., 1969]{Conway1969}
Conway, M.~J., Goodner, C.~J., Werrbach, J.~H., and Gale, C.~C. (1969).
\newblock {Studies of Substrate Regulation in Fasting}.
\newblock {\em The Journal of clinical investigation}, 48:1349--1362.

\bibitem[{De Lorenzo} et~al., 2011]{DeLorenzo2011}
{De Lorenzo}, M.~S., Baljinnyam, E., Vatner, D.~E., Abarz\'{u}a, P., Vatner,
  S.~F., and Rabson, A.~B. (2011).
\newblock {Caloric restriction reduces growth of mammary tumors and
  metastases.}
\newblock {\em Carcinogenesis}, 32(9):1381--7.

\bibitem[Deberardinis et~al., 2008]{DeBerardinis2008}
Deberardinis, R.~J., Sayed, N., Ditsworth, D., and Thompson, C.~B. (2008).
\newblock {Brick by brick: metabolism and tumor cell growth.}
\newblock {\em Current opinion in genetics \& development}, 18(1):54--61.

\bibitem[Demetrakopoulos et~al., 1978]{Demetrakopoulos1978}
Demetrakopoulos, G.~E., Linn, B., and Amos, H. (1978).
\newblock {Rapid loss of ATP by tumor cells deprived of glucose: contrast to
  normal cells}.
\newblock {\em Biochemical and biophysical Research Communications},
  82(3):787--794.

\bibitem[Drenick et~al., 1972]{Drenick1972}
Drenick, E.~J., Alvarez, L.~C., Tamasi, G.~C., and Brickman, A.~S. (1972).
\newblock {Resistance to symptomatic insulin reactions after fasting.}
\newblock {\em The Journal of clinical investigation}, 51(10):2757--62.

\bibitem[Elliott and Barnett, 2011]{Elliott2011}
Elliott, R. and Barnett, B. (2011).
\newblock {Ultrastructural Observation of Mitochondria in Human Breast
  Carcinoma Cells}.
\newblock {\em Microscopy and Microanalysis}, 17(Suppl 2):194--195.

\bibitem[Fine et~al., 2008]{Fine2008}
Fine, E., Segal-Isaacson, C., Feinman, R., and Sparano, J. (2008).
\newblock {Carbohydrate restriction in patients with advanced cancer: a
  protocol to assess safety and feasibility with an accompanying hypothesis}.
\newblock {\em Community Oncology}, 5(1):22--26.

\bibitem[Fine et~al., 2012]{Fine2012}
Fine, E.~J., Segal-Isaacson, C.~J., Feinman, R.~D., Herszkopf, S., Romano,
  M.~C., Tomuta, N., Bontempo, A.~F., Negassa, A., and Sparano, J.~A. (2012).
\newblock {Targeting insulin inhibition as a metabolic therapy in advanced
  cancer: a pilot safety and feasibility dietary trial in 10 patients.}
\newblock {\em Nutrition}, 28(10):1028--35.

\bibitem[Fine et~al., 2009]{Fine2009}
Fine, R.~J., Miller, A., Quadros, E.~V., Sequeira, J.~M., and Feinman, R.~D.
  (2009).
\newblock {Acetoacetate reduces growth and ATP concentration in cancer cell
  lines which over-express uncoupling protein 2.}
\newblock {\em Cancer cell international}, 9(14):1--11.

\bibitem[Frier et~al., 2011]{Frier2011}
Frier, B.~M., Schernthaner, G., and Heller, S.~R. (2011).
\newblock {Hypoglycemia and cardiovascular risks.}
\newblock {\em Diabetes care}, 34 Suppl 2:S132--S137.

\bibitem[Ganapathy-Kanniappan et~al., 2010]{Ganapathy-Kanniappan2010}
Ganapathy-Kanniappan, S., Vali, M., Kunjithapatham, R., Buijs, M., Syed, L.~H.,
  Rao, P.~P., Ota, S., Kwak, B.~K., Loffroy, R., and Geschwind, J.~F. (2010).
\newblock {3-Bromopyruvate: a New Targeted Antiglycolytic Agent and a Promise
  for Cancer Therapy.}
\newblock {\em Current pharmaceutical biotechnology}, 11(5):510--7.

\bibitem[Graham et~al., 2012]{Graham2012}
Graham, N.~A., Tahmasian, M., Kohli, B., Komisopoulou, E., Zhu, M., Vivanco,
  I., Teitell, M.~A., Wu, H., Ribas, A., Lo, R.~S., Mellinghoff, I.~K.,
  Mischel, P.~S., and Graeber, T.~G. (2012).
\newblock {Glucose deprivation activates a metabolic and signaling
  amplification loop leading to cell death.}
\newblock {\em Molecular systems biology}, 8(589):1--16.

\bibitem[Gullino et~al., 1967]{Gullino1967}
Gullino, P., Grantham, F., Courtney, A., and Losonczy, I. (1967).
\newblock {Relationship between oxygen and glucose consumption by transplanted
  tumors in vivo}.
\newblock {\em Cancer research}, 27:1041--1052.

\bibitem[Holm et~al., 1995]{Holm1995}
Holm, E., Hagm\"{u}ller, E., Staedt, U., Schlickeiser, G., Gunther, H.~J.,
  Leweling, H., Tokus, M., and Kollmar, H.~B. (1995).
\newblock {Substrate balances across colonic carcinomas in humans}.
\newblock {\em Cancer research}, 55:1373--1378.

\bibitem[Husain et~al., 2013]{Husain2013}
Husain, Z., Seth, P., and Sukhatme, V.~P. (2013).
\newblock {Tumor-derived lactate and myeloid-derived suppressor cells: Linking
  metabolism to cancer immunology.}
\newblock {\em Oncoimmunology}, 2(11):e26383.

\bibitem[Iqbal et~al., 2013]{Iqbal2013}
Iqbal, M.~A., Siddiqui, F.~A., Gupta, V., Chattopadhyay, S., Gopinath, P.,
  Kumar, B., Manvati, S., Chaman, N., and Bamezai, R. N.~K. (2013).
\newblock {Insulin enhances metabolic capacities of cancer cells by dual
  regulation of glycolytic enzyme pyruvate kinase M2.}
\newblock {\em Molecular cancer}, 12(1):72.

\bibitem[Jalving et~al., 2010]{Jalving2010}
Jalving, M., Gietema, J.~A., Lefrandt, J.~D., de~Jong, S., Reyners, A. K.~L.,
  Gans, R. O.~B., and de~Vries, E. G.~E. (2010).
\newblock {Metformin: taking away the candy for cancer?}
\newblock {\em European journal of cancer}, 46(13):2369--2380.

\bibitem[Jelluma et~al., 2006]{Jelluma2006}
Jelluma, N., Yang, X., Stokoe, D., Evan, G.~I., Dansen, T.~B., and Haas-Kogan,
  D.~A. (2006).
\newblock {Glucose withdrawal induces oxidative stress followed by apoptosis in
  glioblastoma cells but not in normal human astrocytes.}
\newblock {\em Molecular cancer research}, 4(5):319--330.

\bibitem[Jiang and Wang, 2013]{Jiang2013}
Jiang, Y. and Wang, F. (2013).
\newblock {Caloric restriction reduces edema and prolongs survival in a mouse
  glioma model.}
\newblock {\em Journal of neurooncology}, 114(1):25--32.

\bibitem[Jose et~al., 2011]{Jose2011}
Jose, C., Bellance, N., and Rossignol, R. (2011).
\newblock {Choosing between glycolysis and oxidative phosphorylation: a tumor's
  dilemma?}
\newblock {\em Biochimica et biophysica acta}, 1807(6):552--561.

\bibitem[Kim et~al., 1978]{Kim1978}
Kim, S.~H., Kim, J.~H., and Hahn, E.~W. (1978).
\newblock {Selective potentiation of hyperthermic killing of hypoxic cells by
  5-thio-D-glucose}.
\newblock {\em Cancer research}, 38:2935--2938.

\bibitem[Klement and Champ, 2014]{Klement2014}
Klement, R.~J. and Champ, C.~E. (2014).
\newblock {Calories, carbohydrates, and cancer therapy with radiation:
  exploiting the five R's through dietary manipulation.}
\newblock {\em Cancer metastasis reviews}, pages 217--229.

\bibitem[Koppenol et~al., 2011]{Koppenol2011}
Koppenol, W.~H., Bounds, P.~L., and Dang, C.~V. (2011).
\newblock {Otto Warburg's contributions to current concepts of cancer
  metabolism.}
\newblock {\em Nature reviews. Cancer}, 11(5):325--337.

\bibitem[Kossoff et~al., 2007]{Kossoff2007}
Kossoff, E.~H., Turner, Z., and Bergey, G.~K. (2007).
\newblock {Home-guided use of the ketogenic diet in a patient for more than 20
  years.}
\newblock {\em Pediatric neurology}, 36(6):424--425.

\bibitem[Lee et~al., 2012]{Lee2012}
Lee, C., Raffaghello, L., Brandhorst, D., Safdie, F.~M., Bianchi, G.,
  Martin-Montalvo, A., Pistoia, V., Wei, M., Hwang, D., Merlino, S., Emionite,
  L., de~Cabo, T., and Longo, V.~D. (2012).
\newblock {Fasting cycles retard growth of tumors and sensitize a range of
  cancer cell types to chemotherapy.}
\newblock {\em Science translational medicine}, 4(124):124ra27.

\bibitem[Li et~al., 2010]{Li2010a}
Li, Y., Liu, L., and Tollefsbol, T.~O. (2010).
\newblock {Glucose restriction can extend normal cell lifespan and impair
  precancerous cell growth through epigenetic control of hTERT and p16
  expression.}
\newblock {\em FASEB journal}, 24(5):1442--1453.

\bibitem[Magee et~al., 1979]{Magee1979}
Magee, B.~A., Potezny, N., Rofe, A.~M., and Conyers, R.~A. (1979).
\newblock {The inhibition of malignant cell growth by ketone bodies.}
\newblock {\em Australian Journal of Experimental Biology and Medical Science},
  57(5):529--39.

\bibitem[Majumdar et~al., 2009]{Majumdar2009}
Majumdar, A. P.~N., Banerjee, S., Nautiyal, J., Patel, B.~B., Patel, V., Du,
  J., Yu, Y., Elliott, A.~A., Levi, E., and Sarkar, F.~H. (2009).
\newblock {Curcumin synergizes with resveratrol to inhibit colon cancer.}
\newblock {\em Nutrition and cancer}, 61(4):544--53.

\bibitem[Martinez-Outschoorn et~al., 2011]{Martinez-Outschoorn2011}
Martinez-Outschoorn, U.~E., Prisco, M., Ertel, A., Tsirigos, A., Lin, Z.,
  Pavlides, S., Wang, C., Flomenberg, N., Knudsen, E.~S., Howell, A., Pestell,
  R.~G., Sotgia, F., and Lisanti, M.~P. (2011).
\newblock {Ketones and lactate increase cancer cell "stemness," driving
  recurrence, metastasis and poor clinical outcome in breast cancer: achieving
  personalized medicine via Metabolo-Genomics.}
\newblock {\em Cell cycle}, 10(8):1271--86.

\bibitem[Masko et~al., 2010]{Masko2010}
Masko, E.~M., Thomas, J.~A., Antonelli, J.~A., Lloyd, J.~C., Phillips, T.~E.,
  Poulton, S.~H., Dewhirst, M.~W., Pizzo, S.~V., and Freedland, S.~J. (2010).
\newblock {Low-carbohydrate diets and prostate cancer: how low is "low
  enough"?}
\newblock {\em Cancer prevention research}, 3(9):1124--1131.

\bibitem[Mavropoulos et~al., 2009]{Mavropoulos2009}
Mavropoulos, J.~C., Buschemeyer, W.~C., Tewari, A.~K., Rokhfeld, D., Pollak,
  M., Zhao, Y., Febbo, P.~G., Cohen, P., Hwang, D., Devi, G.,
  Demark-Wahnefried, W., Westman, E.~C., Peterson, B.~L., Pizzo, S.~V., and
  Freedland, S.~J. (2009).
\newblock {The effects of varying dietary carbohydrate and fat content on
  survival in a murine LNCaP prostate cancer xenograft model.}
\newblock {\em Cancer prevention research}, 2(6):557--65.

\bibitem[Mavropoulos et~al., 2006]{Mavropoulos2006}
Mavropoulos, J.~C., Isaacs, W.~B., Pizzo, S.~V., and Freedland, S.~J. (2006).
\newblock {Is there a role for a low-carbohydrate ketogenic diet in the
  management of prostate cancer?}
\newblock {\em Urology}, 68(1):15--18.

\bibitem[McCall et~al., 1986]{McCall1986}
McCall, A.~L., Fixman, L.~B., Fleming, N., Tornheim, K., Chick, W., and
  Ruderman, N.~B. (1986).
\newblock {Chronic hypoglycemia increases brain glucose transport.}
\newblock {\em The American journal of physiology}, 251:E442--E447.

\bibitem[McDonald, 1998]{McDonald1998}
McDonald, L. (1998).
\newblock {\em The ketogenic diet: A complete guide for the dieter and
  practitioner}.
\newblock Morris Publishing, Austin, TX.

\bibitem[Moreno-S\'{a}nchez et~al., 2007]{Moreno-Sanchez2007}
Moreno-S\'{a}nchez, R., Rodr\'{\i}guez-Enr\'{\i}quez, S.,
  Mar\'{\i}n-Hern\'{a}ndez, A., and Saavedra, E. (2007).
\newblock {Energy metabolism in tumor cells.}
\newblock {\em The FEBS journal}, 274(6):1393--418.

\bibitem[Morris, 2005]{Morris2005}
Morris, A. A.~M. (2005).
\newblock {Cerebral ketone body metabolism.}
\newblock {\em Journal of inherited metabolic disease}, 28(2):109--21.

\bibitem[Mukherjee et~al., 2004]{Mukherjee2004}
Mukherjee, P., Abate, L.~E., and Seyfried, T.~N. (2004).
\newblock {Antiangiogenic and proapoptotic effects of dietary restriction on
  experimental mouse and human brain tumors}.
\newblock {\em Clinical Cancer Research}, 10:5622--5629.

\bibitem[Mukherjee et~al., 2002]{Mukherjee2002}
Mukherjee, P., El-Abbadi, M.~M., Kasperzyk, J.~L., Ranes, M.~K., and Seyfried,
  T.~N. (2002).
\newblock {Dietary restriction reduces angiogenesis and growth in an orthotopic
  mouse brain tumour model}.
\newblock {\em British Journal of Cancer}, 86:1615--1621.

\bibitem[{National Research Council (US) Food \& Nutrition Board},
  1989]{NRC1989}
{National Research Council (US) Food \& Nutrition Board} (1989).
\newblock {\em Recommended dietary allowances}.
\newblock National Academy Press.

\bibitem[Neal et~al., 2008]{Neal2008}
Neal, E.~G., Chaffe, H., Schwartz, R.~H., Lawson, M.~S., Edwards, N.,
  Fitzsimmons, G., Whitney, A., and Cross, J.~H. (2008).
\newblock {The ketogenic diet for the treatment of childhood epilepsy: a
  randomised controlled trial.}
\newblock {\em Lancet neurology}, 7(6):500--6.

\bibitem[{New York Times}, 1887]{NYT1887}
{New York Times} (1887).
\newblock Sugar and cancer.
\newblock Dec 24.

\bibitem[Niakan, 2010]{Niakan2010}
Niakan, B. (2010).
\newblock {Spontaneous remission of cancer: steady and aggressive malignant
  growth faced with hypoxia or hypoglycemia.}
\newblock {\em Medical hypotheses}, 75(6):505--6.

\bibitem[Oleksyszyn, 2011]{Oleksyszyn2011}
Oleksyszyn, J. (2011).
\newblock {The complete control of glucose level utilizing the composition of
  ketogenic diet with the gluconeogenesis inhibitor, the anti-diabetic drug
  metformin, as a potential anti-cancer therapy.}
\newblock {\em Medical hypotheses}, 77(2):171--3.

\bibitem[Oleksyszyn et~al., 2014]{Oleksyszyn2014}
Oleksyszyn, J., Wietrzyk, J., and Psurski, M. (2014).
\newblock {Cancer – Could it be Cured? A Spontaneous Regression of Cancer,
  Cancer Energy Metabolism, Hyperglycemia-Hypoglycemia, Metformin, Warburg and
  Crabtree Effects and a New Perspective in Cancer Treatment}.
\newblock {\em Journal of Cancer Science \& Therapy}, 06(03):56--61.

\bibitem[Otto et~al., 2008]{Otto2008}
Otto, C., Kaemmerer, U., Illert, B., Muehling, B., Pfetzer, N., Wittig, R.,
  Voelker, H.~U., Thiede, A., and Coy, J.~F. (2008).
\newblock {Growth of human gastric cancer cells in nude mice is delayed by a
  ketogenic diet supplemented with omega-3 fatty acids and medium-chain
  triglycerides.}
\newblock {\em BMC cancer}, 8:122.

\bibitem[Owen et~al., 1967]{Owen1967}
Owen, O.~E., Morgan, A.~P., Kemp, H.~G., Sullivan, J.~M., Herrera, M.~G., and
  Cahill, G.~F. (1967).
\newblock {Brain metabolism during fasting.}
\newblock {\em The Journal of clinical investigation}, 46(10):1589--95.

\bibitem[Pedersen, 1977]{Pedersen1977}
Pedersen, P.~L. (1977).
\newblock Tumor mitochondria and the bioenergetics of cancer cells.
\newblock {\em Progress in experimental tumor research}, 22:190--274.

\bibitem[Phelps, 2004]{Phelps2004}
Phelps, M.~E. (2004).
\newblock {\em {PET}: molecular imaging and its biological applications}.
\newblock Springer.

\bibitem[Phinney et~al., 1983]{Phinney1983}
Phinney, S.~D., Bistrian, B.~R., Wolfe, R.~R., and Blackburn, G.~L. (1983).
\newblock {The Human Metabolic Response to Chronic Ketosis Without Caloric
  Restriction: Physical and Biochemical Adaptation}.
\newblock {\em Metabolism}, 32(8):757--768.

\bibitem[Poff et~al., 2013]{Poff2013}
Poff, A.~M., Ari, C., Seyfried, T.~N., and D'Agostino, D.~P. (2013).
\newblock {The ketogenic diet and hyperbaric oxygen therapy prolong survival in
  mice with systemic metastatic cancer.}
\newblock {\em PloS one}, 8(6):e65522.

\bibitem[Priebe et~al., 2011]{Priebe2011}
Priebe, A., Tan, L., Wahl, H., Kueck, A., He, G., Kwok, R., Opipari, A., and
  Liu, J.~R. (2011).
\newblock {Glucose deprivation activates AMPK and induces cell death through
  modulation of Akt in ovarian cancer cells.}
\newblock {\em Gynecologic oncology}, 122(2):389--95.

\bibitem[Robinson and Williamson, 1980]{Robinson1980}
Robinson, A.~M. and Williamson, D.~H. (1980).
\newblock {Physiological roles of ketone bodies as substrates and signals in
  mammalian tissues.}
\newblock {\em Physiological reviews}, 60(1):143--87.

\bibitem[Sawai et~al., 2004]{Sawai2004}
Sawai, M., Yashiro, M., Nishiguchi, Y., Ohira, M., and Hirakawa, K. (2004).
\newblock {Growth-inhibitory effects of the ketone body, monoacetoacetin, on
  human gastric cancer cells with succinyl-CoA: 3-oxoacid CoA-transferase
  (SCOT) deficiency.}
\newblock {\em Anticancer research}, 24(4):2213--7.

\bibitem[Schmidt et~al., 2011]{Schmidt2011}
Schmidt, M., Pfetzer, N., Schwab, M., Strauss, I., and K\"{a}mmerer, U. (2011).
\newblock {Effects of a ketogenic diet on the quality of life in 16 patients
  with advanced cancer: A pilot trial.}
\newblock {\em Nutrition \& metabolism}, 8(1):54.

\bibitem[Seyfried, 2012]{Seyfried2012}
Seyfried, T. (2012).
\newblock {\em Cancer as a metabolic disease: on the origin, management, and
  prevention of cancer}.
\newblock John Wiley \& Sons.

\bibitem[Seyfried et~al., 2014]{Seyfried2014}
Seyfried, T.~N., Flores, R.~E., Poff, A.~M., and D'Agostino, D.~P. (2014).
\newblock {Cancer as a metabolic disease: implications for novel therapeutics.}
\newblock {\em Carcinogenesis}, 35(3):515--27.

\bibitem[Seyfried et~al., 2008]{Seyfried2008}
Seyfried, T.~N., Kiebish, M., Mukherjee, P., and Marsh, J. (2008).
\newblock {Targeting energy metabolism in brain cancer with calorically
  restricted ketogenic diets.}
\newblock {\em Epilepsia}, 49 Suppl 8:114--6.

\bibitem[Seyfried et~al., 2003]{Seyfried2003}
Seyfried, T.~N., Sanderson, T.~M., El-Abbadi, M.~M., McGowan, R., and
  Mukherjee, P. (2003).
\newblock {Role of glucose and ketone bodies in the metabolic control of
  experimental brain cancer.}
\newblock {\em British journal of cancer}, 89(7):1375--82.

\bibitem[Seyfried and Shelton, 2010]{Seyfried2010}
Seyfried, T.~N. and Shelton, L.~M. (2010).
\newblock {Cancer as a metabolic disease.}
\newblock {\em Nutrition \& metabolism}, 7(7):1----22.

\bibitem[Shelton et~al., 2010]{Shelton2010}
Shelton, L.~M., Huysentruyt, L.~C., Mukherjee, P., and Seyfried, T.~N. (2010).
\newblock {Calorie restriction as an anti-invasive therapy for malignant brain
  cancer in the VM mouse.}
\newblock {\em ASN neuro}, 2(3):e00038.

\bibitem[Simone et~al., 2013]{Simone2013}
Simone, B.~A., Champ, C.~E., Rosenberg, A.~L., Berger, A.~C., Monti, D.~A.,
  Dicker, A.~P., and Simone, N.~L. (2013).
\newblock {Selectively starving cancer cells through dietary manipulation:
  methods and clinical implications}.
\newblock {\em Future Oncology}, 9(7):959--976.

\bibitem[Singh and Lai, 2004]{Singh2004}
Singh, N.~P. and Lai, H.~C. (2004).
\newblock Artemisinin induces apoptosis in human cancer cells.
\newblock {\em Anticancer Research}, 24(4):2277--2280.

\bibitem[Sivananthan, 2013]{Sivananthan2013}
Sivananthan, A.~P. (2013).
\newblock {\em {Effects of a Ketogenic Diet on Tumor Progression in Breast
  Cancer}}.
\newblock PhD thesis, Icahn School of Medicine at Mount Sinai.

\bibitem[Skinner et~al., 2009]{Skinner2009}
Skinner, R., Trujillo, A., Ma, X., and Beierle, E.~A. (2009).
\newblock {Ketone bodies inhibit the viability of human neuroblastoma cells.}
\newblock {\em Journal of pediatric surgery}, 44(1):212--6; discussion 216.

\bibitem[Sokoloff, 1973]{Sokoloff1973}
Sokoloff, L. (1973).
\newblock {Metabolism of ketone bodies by the brain}.
\newblock {\em Annual review of medicine}, 24:271--280.

\bibitem[Spitz et~al., 2000]{Spitz2000}
Spitz, D.~R., Sim, J.~E., Ridnour, L. A.~A., Galoforo, S.~S., and Lee, Y.~J.
  (2000).
\newblock {Glucose deprivation-induced oxidative stress in human tumor cells. A
  fundamental defect in metabolism?}
\newblock {\em Annals of the New York Academy of Sciences}, 899:349--362.

\bibitem[Stafford et~al., 2010]{Stafford2010}
Stafford, P., Abdelwahab, M.~G., Kim, D.~Y., Preul, M.~C., Rho, J.~M., and
  Scheck, A.~C. (2010).
\newblock {The ketogenic diet reverses gene expression patterns and reduces
  reactive oxygen species levels when used as an adjuvant therapy for glioma.}
\newblock {\em Nutrition \& metabolism}, 7:74.

\bibitem[Stewart and Fleming, 1973]{Stewart1973}
Stewart, W.~K. and Fleming, L.~W. (1973).
\newblock {Features of a successful therapeutic fast of 382 days' duration.}
\newblock {\em Postgraduate medical journal}, 49(569):203--9.

\bibitem[Tan-Shalaby and Seyfried, 2013]{TanShalaby2013}
Tan-Shalaby, J. and Seyfried, T. (2013).
\newblock {Ketogenic Diet in Advanced Cancer: A Pilot Feasibility and Safety
  Trial in the Veterans Affairs Cancer Patient Population}.
\newblock {\em Journal of Clinical Trials}, 3(04):4--7.

\bibitem[Tin et~al., 2007]{Tin2007}
Tin, M. M.~Y., Cho, C.-H., Chan, K., James, A.~E., and Ko, J. K.~S. (2007).
\newblock Astragalus saponins induce growth inhibition and apoptosis in human
  colon cancer cells and tumor xenograft.
\newblock {\em Carcinogenesis}, 28(6):1347--1355.

\bibitem[van Poelje et~al., 2011]{Poelje2011}
van Poelje, P.~D., Potter, S.~C., and Erion, M.~D. (2011).
\newblock {Diabetes - Perspectives in Drug Therapy}.
\newblock {\em Handbook of Experimental Pharmacology}, 203:279--301.

\bibitem[Vogelstein et~al., 2013]{Vogelstein2013}
Vogelstein, B., Papadopoulos, N., Velculescu, V.~E., Zhou, S., Diaz, L.~A., and
  Kinzler, K.~W. (2013).
\newblock {Cancer genome landscapes.}
\newblock {\em Science}, 339(6127):1546--58.

\bibitem[Warburg et~al., 1924]{Warburg1924}
Warburg, O., Posener, K., and Negelein, E. (1924).
\newblock Ueber den stoffwechsel der tumoren.
\newblock {\em Biochemische Zeitschrift}, 152(1):319--344.

\bibitem[Waxman et~al., 1995]{Waxman1995}
Waxman, S.~G., Kocsis, J.~D., and Stys, P.~K. (1995).
\newblock {\em The axon: structure, function, and pathophysiology}.
\newblock Oxford University Press.

\bibitem[Woolf and Scheck, 2014]{Woolf2014}
Woolf, E.~C. and Scheck, A.~C. (2014).
\newblock {The Ketogenic Diet for the Treatment of Malignant Glioma.}
\newblock {\em Journal of lipid research}, pages 1--19.

\bibitem[Yuneva, 2008]{Yuneva2008}
Yuneva, M. (2008).
\newblock {Finding an ``Achilles' heel'' of cancer: The role of glucose and
  glutamine metabolism in the survival of transformed cells}.
\newblock {\em Cell Cycle}, 7(14):2083--2089.

\bibitem[Zhou et~al., 2007]{Zhou2007}
Zhou, W., Mukherjee, P., Kiebish, M.~A., Markis, W.~T., Mantis, J.~G., and
  Seyfried, T.~N. (2007).
\newblock {The calorically restricted ketogenic diet, an effective alternative
  therapy for malignant brain cancer.}
\newblock {\em Nutrition \& metabolism}, 4:5.

\bibitem[Zuccoli et~al., 2010]{Zuccoli2010}
Zuccoli, G., Marcello, N., Pisanello, A., Servadei, F., Vaccaro, S., Mukherjee,
  P., and Seyfried, T.~N. (2010).
\newblock {Metabolic management of glioblastoma multiforme using standard
  therapy together with a restricted ketogenic diet: Case Report.}
\newblock {\em Nutrition \& metabolism}, 7(33):1---7.

\end{thebibliography}

\end{document}